\documentclass{aastex7}

\usepackage{hyperref}
\usepackage{orcidlink}
\usepackage{graphicx}
\usepackage{amsmath}

\begin{document}

\title{ On the origin of the X-ray emission surrounding PSR B0656+14 in the eROSITA Cal-PV data }

\author[orcid=0000-0002-0456-7229]{Alena Khokhriakova}
\affiliation{Max-Planck-Institut f\"ur extraterrestrische Physik,
     Giessenbachstra{\ss}e, 85748 Garching, Germany}
\email[show]{alena@mpe.mpg.de}  

\author[orcid=0000-0003-1173-6964]{Werner Becker} 
\affiliation{Max-Planck-Institut f\"ur extraterrestrische Physik,
     Giessenbachstra{\ss}e, 85748 Garching, Germany}
\email{web@mpe.mpg.de}
\affiliation{Max-Planck-Institut f\"ur Radioastronomie,
     Auf dem H\"ugel 69, 53121 Bonn, Germany }

\author[orcid=0000-0002-2708-5097]{Peter Predehl}
\affiliation{Max-Planck-Institut f\"ur extraterrestrische Physik,
     Giessenbachstra{\ss}e, 85748 Garching, Germany}
\email{ }
 
\author[orcid=0000-0003-2189-4501]{Jeremy S. Sanders}
\affiliation{Max-Planck-Institut f\"ur extraterrestrische Physik,
     Giessenbachstra{\ss}e, 85748 Garching, Germany}
\email{ }

\author[orcid=0000-0001-8158-4631]{Michael Freyberg}
\affiliation{Max-Planck-Institut f\"ur extraterrestrische Physik,
     Giessenbachstra{\ss}e, 85748 Garching, Germany}
\email{mjf@mpe.mpg.de}

\author[orcid=0000-0003-3441-9355]{Axel Schwope}
\affiliation{Leibniz-Institut für Astrophysik Potsdam (AIP), An der Sternwarte 16, 14482 Potsdam, Germany}
\email{aschwope@aip.de}

\begin{abstract}

We present a cautionary assessment of the extended X-ray emission around PSR~B0656+14 in eROSITA Cal-PV data in response to the work of \cite{2025arXiv250117046N}.
The eROSITA PSF model is known to underestimate emission in the wings beyond~$1'$. This prevents a reliable detection of faint nebular emission around PSR~B0656+14 as claimed by \cite{2025arXiv250117046N}.
In addition, spectral analysis shows the surrounding diffuse X-rays can be fitted with the same 2BB+PL model as the pulsar's emission itself.
This strongly invalidates the interpretation by  \cite{2025arXiv250117046N} that the X-ray emission in the $(4-10)'$ region is associated with the degree-scale gamma-ray halo recently found by the High-Altitude Water Cherenkov Observatory (HAWC), and shows that it originates from the pulsar due to the wings of the PSF.

\end{abstract}


\section{TeV halos around middle-aged pulsars} 

Recent observations by HAWC, LHAASO, and Fermi-LAT have revealed a class of extended gamma-ray emission surrounding middle-aged pulsars, known as gamma-ray halos or TeV halos \citep{2017Sci...358..911A, 2021PhRvL.126x1103A}.  
These halos are typically a few degrees across (corresponding to tens of parsecs at kiloparsec distances), significantly larger than the much more compact pulsar wind nebulae (PWNe) observed at X-ray and radio wavelengths.  
Their gamma-ray emission is thought to arise from inverse Compton scattering of ambient photon fields -- primarily the cosmic microwave background -- off relativistic electrons and positrons that have escaped the PWN into the surrounding interstellar medium (ISM).

Theoretical studies predict diffuse X-ray synchrotron halos around pulsars, produced by the same electrons responsible for TeV halos via inverse Compton scattering.
For the archetypal Geminga halo, deep XMM‑Newton and NuSTAR observations found no extended emission, yielding strong upper limits on the X-ray flux at degree scales \citep{2024A&A...689A.326M}.
These limits suggest either a very low ambient magnetic field ($\le 1 \mu$G) or low energy of the electrons in the outer halo.

A dedicated eROSITA study of middle-aged pulsars, including Geminga and PSR B0656+14, found no degree-scale X-ray halos, constraining surface brightness levels \citep{2024A&A...683A.180K}.
Similarly, Swift-XRT observations of the TeV halo candidate HESS J1813–126 yielded non-detections, reinforcing the view that TeV halos may not universally produce detectable synchrotron counterparts \citep{2025arXiv250408689G}.

\section{The eROSITA Cal-PV observation of PSR B0656+14}

These upper limits have motivated deeper X-ray searches around pulsars hosting TeV halos. 
One such study by \citet{2025arXiv250117046N} investigated PSR B0656+14 using eROSITA pointed observations from the early ``Calibration and Performance Verification'' (Cal-PV) phase (obsID 300000).
While \citet{2024A&A...683A.180K} found no large-scale diffuse emission in stacked survey data, \citet{2025arXiv250117046N} reported faint X-ray emission of $\sim (4–10)'$ scale around PSR B0656+14 in pointed Cal-PV data.
Notably, such emission was not reported in the original paper describing the Cal-PV observations of B0656+14 by the eROSITA collaboration \citep{2022A&A...661A..41S}.
Although the reported feature is considerably smaller than the degree-wide TeV halo surrounding PSR B0656+14, the authors suggest a potential association. 
Modeling the extended emission with a power law, they find $N_{\text{H}} = 2.1^{+1.4}_{-1.2} \times 10^{20} \text{ cm}^{-2}$ and a power-law photon-index $\Gamma = -3.7 \pm 0.4$, assuming $N_{\text{H}}$ in the range $(1-2.8) \times 10^{20} \text{ cm}^{-2}$. 

\section{Limitations of the eROSITA Point Spread Function}

The reported extended X-ray emission, separate from the pulsar emission, interpreted as a smaller counterpart of the TeV halo, appears questionable due to limitations in eROSITA’s PSF modeling in pointed mode. 
The PSF was characterized using pre-launch PANTER data and in-flight observations.
PANTER measurements, which extend to $4'$, were used to construct a shapelet-based analytical PSF model \citep{2022A&A...661A...1B}.
This model reproduces PANTER enclosed fluxes within $(10-20)$\% accuracy, which is sufficient for many applications.

The shapelet model uses a maximum scale of 6 pixels ($\sim1'$), limiting its accuracy beyond this radius.
This limitation is evident in Fig.~A.1 and Appendix~A of \citet{2024A&A...682A..34M}, which show clear model–data discrepancies at large radii. 
Similarly,  Fig.~A.1 of \citet{2023A&A...670A.156C} shows that significant flux extends beyond $4'$.
While the shapelet model can still match the enclosed energy fraction up to $\sim 4'$ in an average sense, it does so without explicitly modeling the spatial distribution of the PSF wings beyond $1'$. 
Therefore, its predictive power for the surface brightness profile at larger angular scales is inherently limited.
In their analysis, the authors of \citet{2025arXiv250117046N} applied this shapelet-based model to estimate the contribution of the pulsar PSF at radii $> 4'$. However, given the modeling limitations discussed above, such extrapolation must be treated with great caution. The diffuse X-ray emission they report in this range may arise from unmodeled PSF wings rather than representing a physically distinct component. 

\section{Independent spectral re-analysis}

\begin{figure*}[ht!]
\includegraphics[width=\textwidth]{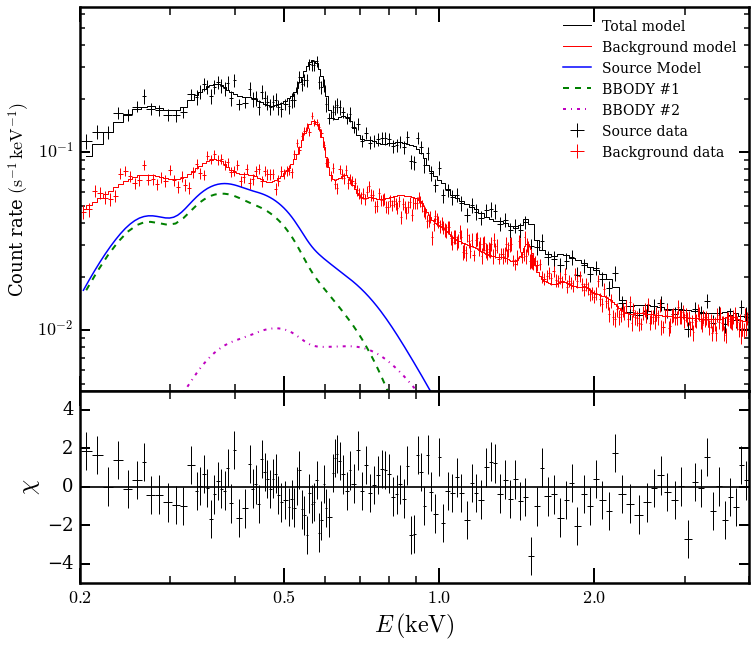}
\caption{ The X-ray spectrum of the diffuse emission in the $(4-6)'$ annulus around PSR~B0656+14 fitted with the pulsar's spectral model, consisting of two blackbody components and a power-law. The solid blue line shows the total source model, while the dashed green and magenta dash-dot lines indicate the first and second blackbody components, respectively. The power-law component is not shown, as its contribution is negligible; only an upper limit on its normalization could be derived.
\label{fig:spectrum}}
\end{figure*}

To test the origin of the reported diffuse emission, we independently re-analyzed the same regions as in \cite{2025arXiv250117046N}. 
The authors report the results of a spectral analysis, modeling the diffuse emission with a simple power-law having a slope $\Gamma \approx-3.7$. 
This unusually soft spectrum differs from TeV halo properties and suggests the emission could instead match the pulsar’s thermal model.
We first tested  the same spectral model as in \cite{2025arXiv250117046N}. 
We confirmed their result, finding  $\Gamma = -3.51^{+0.13}_{-0.03}$ with a goodness of fit of 1.06. 
We then applied the pulsar model from \cite{2022A&A...661A..41S}, based on eROSITA Cal-PV data. 
This model includes two blackbody components plus a power law. 
Fixing temperatures and photon index to their best-fit values, we obtained the following normalizations: $K_{\text{BB1}} = 250^{+23}_{-21}$, $K_{\text{BB2}} = 2.9^{+1.5}_{-1.3}$, $K_{\text{PL}} < 2.8 \times 10^{-7}$ keV$^{-1}$ cm$^{-2}$ s$^{-1}$; goodness of fit 1.08. 
All normalizations are smaller due to reduced number of photons in the $(4-10)'$ annulus, and the non-thermal component is not statistically required.
The goodness of fit is comparable to that of the power-law model in \cite{2025arXiv250117046N}. 
This shows that the diffuse emission in the $(4-10)'$ surroundings of the pulsar can be modeled with the same spectral model as the pulsar emission itself.
Fig.~\ref{fig:spectrum} shows the $(4-6)'$ spectrum and best-fit model.

Our results suggest the signal in the $(4-10)'$ annulus can be fully explained by PSF leakage from the pulsar, rather than a separate diffuse structure. 
Given the current limitations in characterizing the eROSITA pointed-mode PSF beyond $\sim 1'$, we caution against interpreting similar extended features as astrophysical without further validation. 

So far, eROSITA has only performed pointed observations during its Cal-PV phase and following orbital correction maneuvers. These observations were primarily intended to test the observatory and instrument performance in space, rather than to provide calibrated science data. 
Since February 2022, the mission has been in safe mode, and no new pointed observations can currently be acquired. 
While future PSF improvements are possible, the current in-flight PSF characterization remains limited, particularly for pointed-mode observations. This is critical for Cal-PV data, where a detailed understanding of instrument response and careful treatment of calibration uncertainties are essential. Therefore, analyses based on current PSF models must be interpreted cautiously, as inaccuracies in the PSF wings can significantly affect studies of faint diffuse emission around bright point sources.

\facilities{eROSITA}

\software{
          XSPEC \citep{Arnaud1996}
          }

\bibliography{bib}{}
\bibliographystyle{aasjournalv7}

\end{document}